\documentclass[floatfix,%
 aip,
 amsmath,amssymb,
 reprint,%
]{revtex4-1}
\usepackage{placeins}
\usepackage[]{amsmath}
\usepackage{amssymb}

\usepackage{graphicx}
\usepackage{dcolumn}
\usepackage{bm}
\usepackage[utf8]{inputenc}
\usepackage[T1]{fontenc}
\usepackage{mathptmx}
\usepackage{etoolbox}
\usepackage[]{color}
\usepackage{ulem}
\usepackage{hyperref}

\usepackage{times,psfrag,subfigure}
\usepackage{syntonly}
\usepackage{lmodern}
\usepackage[version=4]{mhchem}
\usepackage{siunitx}
\makeatletter
\def\@email#1#2{%
 \endgroup
 \patchcmd{\titleblock@produce}
  {\frontmatter@RRAPformat}
  {\frontmatter@RRAPformat{\produce@RRAP{*#1\href{mailto:#2}{#2}}}\frontmatter@RRAPformat}
  {}{}
}%
\makeatother
\begin{document}

\preprint{AIP/123-QED}

\title{Interfacially arrested melting in thin films: capillarity-driven suspension of phase transitions}

\author{Chenyu Jin}
\email{chenyujin@ywfudan.cn}
\affiliation{ 
  Zhejiang Key Laboratory of Extreme Environment Functional Materials, Yiwu Research Institute of Fudan University, Zhejiang 322000, People's Republic of China
}

\author{Guoxiang Chen}
\affiliation{ 
  Zhejiang Key Laboratory of Extreme Environment Functional Materials, Yiwu Research Institute of Fudan University, Zhejiang 322000, People's Republic of China
}

\author{Beibei Wang}
\affiliation{ 
  Zhejiang Key Laboratory of Extreme Environment Functional Materials, Yiwu Research Institute of Fudan University, Zhejiang 322000, People's Republic of China
}

\author{Yongfeng Mei}
\affiliation{ 
  Zhejiang Key Laboratory of Extreme Environment Functional Materials, Yiwu Research Institute of Fudan University, Zhejiang 322000, People's Republic of China
}
\affiliation{ 
  State Key Laboratory of Surface Physics $\&$ International Institute for Intelligent Nanorobots and Nanosystems, College of Intelligent Robotics and Advanced Manufacturing, Fudan University, Shanghai 200438, People's Republic of China
}
\affiliation{ 
  Shanghai Frontiers Science Research Base of Intelligent Optoelectronics and Perception, Institute of Optoelectronics, Fudan University, Shanghai 200438, People's Republic of China
}

\author{Hans Riegler}
\affiliation{Max Planck Institute of Colloids and Interfaces, Science Park Golm, 14476 Potsdam, Germany}
\date{\today}

\begin{abstract}
Melting is typically viewed as a bulk first-order phase transition that proceeds once nucleation barriers are overcome. Here we demonstrate an interfacially arrested melting regime in molecularly thin crystalline films, where large liquid droplets remain stably trapped well above the bulk melting temperature. Using long-chain alkane films as a model system, we show that melting is suspended by the competition between bulk melting enthalpy and interfacial energy costs associated with capillary confinement. The arrested state is governed by a single control parameter, the product of the temperature offset and the embedding depth, and is independent of droplet size. As a consequence, small temperature variations produce pronounced and reversible changes in droplet morphology, enabling intrinsic thermodynamic amplification of thermal signals. These results reveal a general mechanism by which interfacial constraints can arrest first-order phase transitions in thin films.


\end{abstract}
\maketitle


Melting, as a first-order phase transition, is generally expected to proceed once nucleation barriers are overcome, with the liquid phase rapidly growing until the solid disappears \cite{adkins1983equilTD, kuhlmann-wilsdorf_theory_1965, rettenmayr_solidification_2009, de_with_melting_2023}. Stable coexistence of solid and liquid far from the bulk melting temperature is therefore usually associated with kinetic limitations, hysteresis, or external constraints \cite{holmberg_phase_2009, han_phase_2010, agrawal_observation_2017, yao_designing_2021}. However, recent studies have shown that in confined or interfacially complex systems, local energetic costs can strongly modify phase-transition pathways \cite{dash_physics_2006, moerz2012capillary, huber2015soft, christenson2013two, kusumaatmaja2012bulge}. Whether melting can be thermodynamically arrested after nucleation remains an open question. This question is broadly relevant to phase transitions in micro- and nanoscale systems, where confinement and interfacial energies often become comparable to bulk thermodynamic driving forces \cite{Glicksman2011, tartaglino_melting_2005}. In such regimes, phase transitions can deviate strongly from their bulk behavior, with important consequences for thin-film stability, pattern evolution, and process control in micro- and nano-fabrication \cite{hammoud_pressureless_2025, visser2015toward, sahu2017shaping}.

Thin crystalline films provide a natural platform to explore this possibility, as melting in such systems inevitably involves the creation and deformation of interfaces. In particular, when lateral advance of a melting front is geometrically or topologically constrained, bulk melting enthalpy may compete directly with interfacial energy costs \cite{LuKe2007rev, zhang2000superheating}. Long-chain alkane films offer an ideal model system in this context: they exhibit surface freezing rather than surface premelting, partial wetting of the solid by the melt, and molecularly thin crystalline layers with well-defined thickness and heterogeneous crystalline facets \cite{Merkl1997SF, riegler2007pre, pithan2015}. These features make it possible to isolate interfacial contributions to melting while minimizing kinetic effects \cite{Lazar2005movingdrop, kusumaatmaja2012bulge, jin2016island, jin2023hole}.
Previously, we have used long-chain alkane systems with excess material beyond a monolayer to study melting pathways of three-dimensional aggregates, where the vertical dimension of the structures extends far beyond that of molecularly thin films \cite{jin2016island}.

In the present work, although multilayer structures are also present, the system remains in the thin-film limit, where melting is governed by interfacial confinement rather than bulk geometry. Here we demonstrate an interfacially arrested melting regime in molecularly thin alkane films containing laterally confined liquid droplets. We show that large liquid droplets can coexist stably with crystalline solid layers well above the bulk melting temperature, without kinetic hysteresis. Melting is suspended by the balance between bulk melting enthalpy and interfacial energy associated with capillary confinement. This balance is governed by a single thermodynamic control parameter, the product of the temperature offset and the embedding depth, and is independent of droplet size. As a consequence, small temperature variations induce pronounced and reversible changes in droplet morphology, revealing an intrinsic thermodynamic amplification of thermal signals. These results identify a general mechanism by which interfacial constraints can arrest first-order phase transitions in thin films.


\begin{figure*}
  \centering
  \includegraphics[width=.8\textwidth]{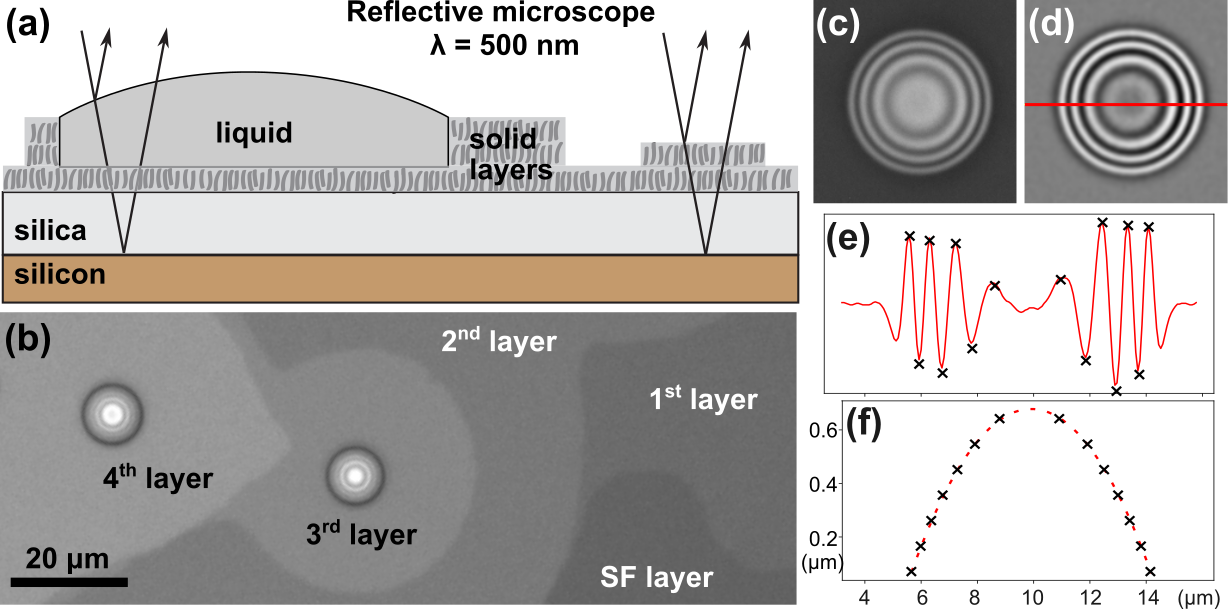}
  \caption{Schematic illustration of the experimental setup and data analysis procedure. Optical interference between light reflected from the alkane–air and silicon–silica interfaces enhances the vertical resolution (a), providing sufficient contrast to resolve single alkane layer with a thickness of approximately $4\,\mathrm{nm}$ (b). To extract droplet profiles, raw image (c) is first processed using a high-pass filter (d). The positions of the peaks and valleys of the resulting Newton rings are then identified and converted into height information (e). Owing to the static and axisymmetric nature of the droplets, the reconstructed profiles are fitted with a circular cap to resolve the apparent contact angle (f).
} 
  \label{fig:setup}
\end{figure*}

Planar silica substrates were coated with an excess monolayer of triacontane ($\ce{C_{30}H_{62}}$). Long-chain alkanes form a surface-frozen (SF) monolayer, i.e., a crystalline monolayer between liquid and wafer surface, slightly below the bulk melting temperature $T_0$, which acts as the effective substrate in our experiments. Excess material segregates into liquid droplets and molecularly thin crystalline films composed of upright-oriented alkane molecules, yielding a single-layer thickness of approximately $4\,\mathrm{nm}$ consistent with atomic force microscopy measurements and the molecular all-trans length. Temperature was controlled with an accuracy of $0.1^\circ$C.

The thickness of the films and the morphology of the droplets were resolved using optical interference microscopy. A specially prepared substrate -- a silicon wafer with a $300$-nm-thick thermal oxide layer -- was employed to achieve nanometer-scale vertical resolution. This enhancement arises from optical interference: the refractive index of silica ($n = 1.46$) closely matches that of $n$-alkane ($n = 1.45$ for liquid and $n = 1.50$ for solid), while silicon has a much higher refractive index ($n = 3.88$). As a result, incident light is primarily reflected from the alkane–air and silicon–silica interfaces (Fig.~\ref{fig:setup}(a)). The oxide thickness was chosen to optimize contrast for thin alkane layers under green illumination \cite{kohler2006optical}.  

As shown in Fig.~\ref{fig:setup}(b), molecularly thin films can be clearly distinguished, and their thickness resolved by counting discrete layers. Beyond the surface-freezing (SF) layer, additional solid layers grow slowly (minutes to hours) via stochastic nucleation \cite{Lazar2005flow}, leading to non-overlapping terraces with spatially varying interference contrast (Fig.~\ref{fig:setup}(b)). These contrast variations reveal the topography of the solid landscape, allowing us to resolve the number of solid layers beneath or around a given droplet, and thus infer its embedding state. In the following, we focus on the confined-droplet configuration with a single surface-frozen (SF) layer beneath the liquid.

Simultaneously, Newton rings were used to extract droplet profiles and apparent contact angles by fitting the interference pattern (Fig.~\ref{fig:setup}(c)--(f)). We define the apparent contact angle, $\theta_a$, as the macroscopic angle formed by the liquid interface with respect to the substrate, as measured from optical images. It characterizes the effective geometry of the droplet at the optical scale and may differ from the Young contact angle due to the presence of a nanometric surface-frozen layer. In addition, confined droplets can be distinguished from reference droplets by their distinct response to temperature changes.

Further experimental details are provided in the Supporting Information.


Due to the chain-like structure of $n$-alkanes, the solid film is organized in discrete molecular layers. The top surface of the solid film is a (001) chain-end facet, while the film sidewalls correspond to higher-energy side facets parallel to the molecular axis. Liquid alkane only partially wets the top surface, with a contact angle of about $20^\circ$. Phase transition at this interface is kinetically trapped -- both melting and freezing need to overcome a nucleation barrier. Hence, a droplet resting on a solid thin film may coexist with solid over several degrees around bulk melting temperature $T_0$.

\begin{figure*}
  \centering
  \includegraphics[width=.8\textwidth]{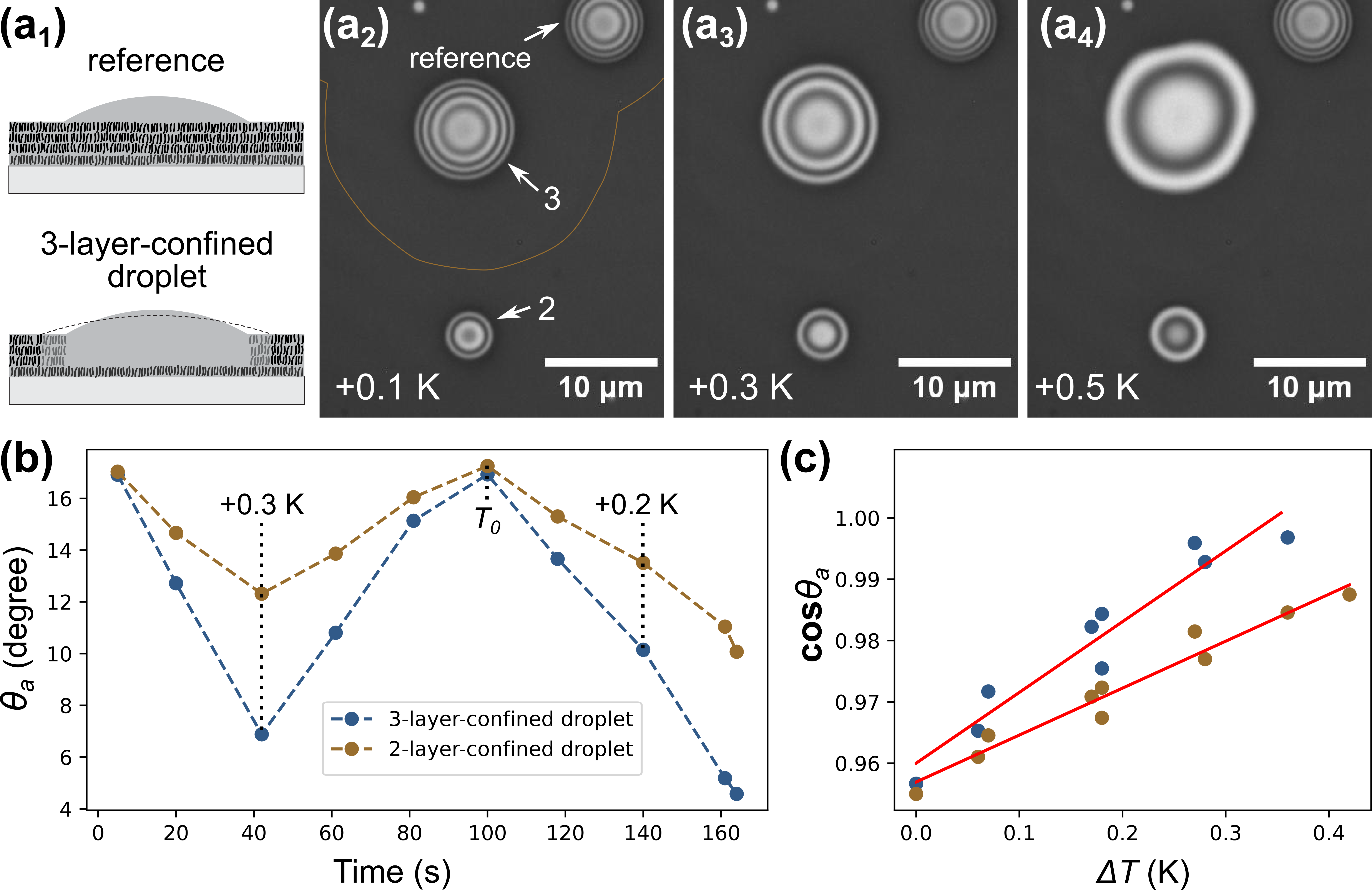}
  \caption{Reversible expansion and retraction of droplets with temperature. (a) Two droplets embedded in alkane films flatten as temperature increases. The response is more pronounced for the droplet trapped in a three-layer film (droplet $\#3$) than for that in a two-layer film (droplet $\#2$), whereas a reference droplet resting on a solid film shows no measurable change in the apparent contact angle $\theta_a$. The edge of the three-layer film is marked by a thin line. (b) The morphological response is fully reversible upon temperature cycling. (c) As predicted by the energetic analysis, $\cos\theta_a$ varies linearly with the temperature offset.
}
  \label{fig:ca}
\end{figure*}

During the sample preparation, some liquid droplets can be confined within the thin films. These confined droplets are identified by their characteristic thermal response: Upon increasing temperature above $T_0$, the confined droplets undergo pronounced and fully reversible changes in morphology. Specifically, the droplets flatten and their apparent contact angles decrease. This response is reversible upon cooling, with no detectable hysteresis within the experimental resolution (Fig.~\ref{fig:ca}(a-b)). By contrast, a reference droplet resting on a solid thin film shows no measurable change in contact angle or shape over the same temperature range. These observations indicate that melting proceeds only partially and reaches stationary states governed by a quasi-equilibrium balance between bulk and interfacial energies, rather than by kinetic effects such as contact-line pinning.

A systematic dependence on the embedding depth $h$ is observed. Droplets confined within thicker solid films exhibit a significantly stronger morphological response to temperature changes than those in thinner films. Experimentally, we find that $\cos\theta_a$ varies linearly with the temperature offset $\Delta T= T - T_0$ throughout the suspended melting regime, as shown in Fig.~\ref{fig:ca}(c), and the slope of $\cos\theta_a$ versus $\Delta T$ increases systematically with the embedding depth.

\begin{figure}
  \centering
  \includegraphics[width=\columnwidth]{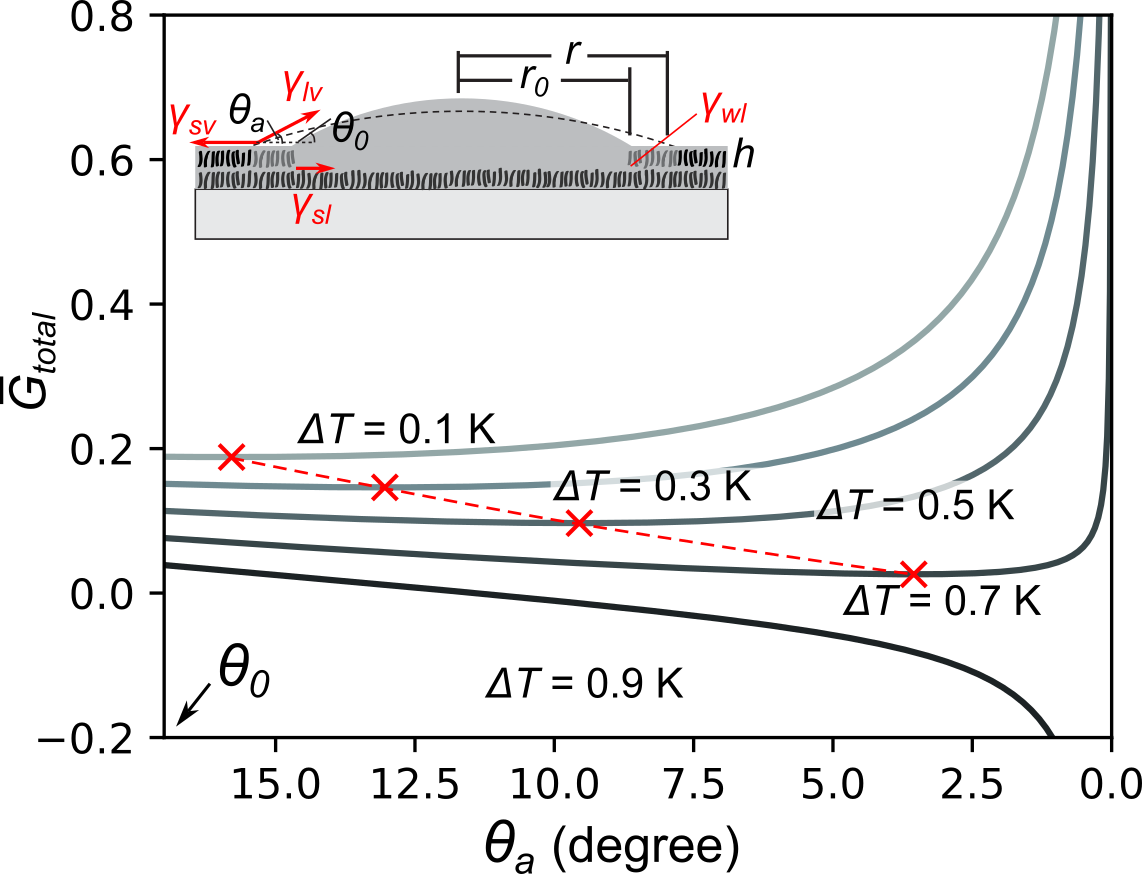}
    \caption{Normalized free energy $\bar{G}_{\mathrm{total}} = G_{\mathrm{total}}/(\gamma_{lv} r_0^2)$ as a function of the apparent contact angle $\theta_a$, with $\gamma_{lv}$ the liquid–vapor interfacial tension and $r_0$ the droplet radius at $T_0$. When melting proceeds with the liquid front pinned at the solid boundary, $\theta_a$ decreases from the Young contact angle $\theta_0=17^\circ$ toward zero. The embedding depth is fixed at $4\,\mathrm{nm}$, and curves correspond to increasing $\Delta T$ from $0.1\,\mathrm{K}$ to $0.9\,\mathrm{K}$.
    In the arrested melting regime, variations in the control parameter $\Delta T \cdot h$ modify the balance between bulk and interfacial energies, leading to a continuous shift of the free-energy minimum (red crosses). Inset: schematic of a confined droplet, illustrating the geometric constraints and relevant parameters of the energetic model.
   }
  \label{fig:drop-G}
\end{figure}

Droplets laterally confined within solid films are in direct contact with the high-energy film sidewalls, where the liquid completely wets the solid (contact angle $0$). Phase transformation at this interface therefore does not involve a nucleation barrier, and melting would ordinarily be expected to proceed once the temperature exceeds the bulk melting temperature $T_0$ \cite{herring1951some, nozieres1989surface}. In the present system, however, the liquid front remains pinned at the edge of the solid film. As a result, melting does not advance laterally but instead enlarges the liquid--vapor interface, appearing macroscopically as a progressive flattening of the droplet (Fig.~\ref{fig:drop-G}, inset).

Because the droplet radius satisfies $r \gg h$, even a small amount of melting leads to a substantial increase in liquid--vapor interfacial area. The associated capillary energy cost can therefore become comparable to the bulk free-energy gain from melting and counterbalance the thermodynamic driving force. This competition arrests further melting and stabilizes a partially melted state.

To elucidate the origin of this arrested melting, we consider the competition between bulk and interfacial contributions to the free energy. 
The total free energy of the system is written as
\[
G_{\mathrm{total}} = G_{\mathrm{bulk}} + G_{\mathrm{interface}},
\]
where $G_{\mathrm{bulk}}$ represents the free-energy gain associated with melting, and $G_{\mathrm{interface}}$ accounts for the energetic cost of creating liquid--vapor and liquid--solid interfaces. 
At equilibrium, these two contributions balance,
\[
-dG_{\mathrm{bulk}} = dG_{\mathrm{interface}} .
\]

The competition between bulk and interfacial contributions generates a free-energy minimum at finite $\theta_a$, corresponding to a stationary, partially melted state (energy minima marked by red dots in Fig.~\ref{fig:drop-G}). This equilibrium is controlled by the single parameter $\Delta T \cdot h$, and is independent of droplet size. With increasing temperature, the free-energy minimum shifts continuously toward smaller $\theta_a$, resulting in reversible changes in droplet morphology without complete melting.

The pinned liquid front further imposes a geometric constraint: the volume of the spherical-cap top of the droplet, $V_c$, remains constant. Under this constraint, changes in the liquid--vapor interfacial area obey
\[
dA_{lv}\big|_{V_c=\mathrm{const}} = 2\pi r \cos\theta_a \, dr .
\]

Combining this relation with the energy balance yields
\begin{equation}
\cos \theta_a =\cos \theta_0 + \frac{\Delta S_\text{fus} \cdot \Delta T \cdot h}{\gamma _{lv}} 
	\label{eqn:cos}
\end{equation}
which governs the arrested melting state, with $\Delta S_\text{fus}$ the entropy of fusion per volume.

\begin{figure}
  \includegraphics[width=\columnwidth]{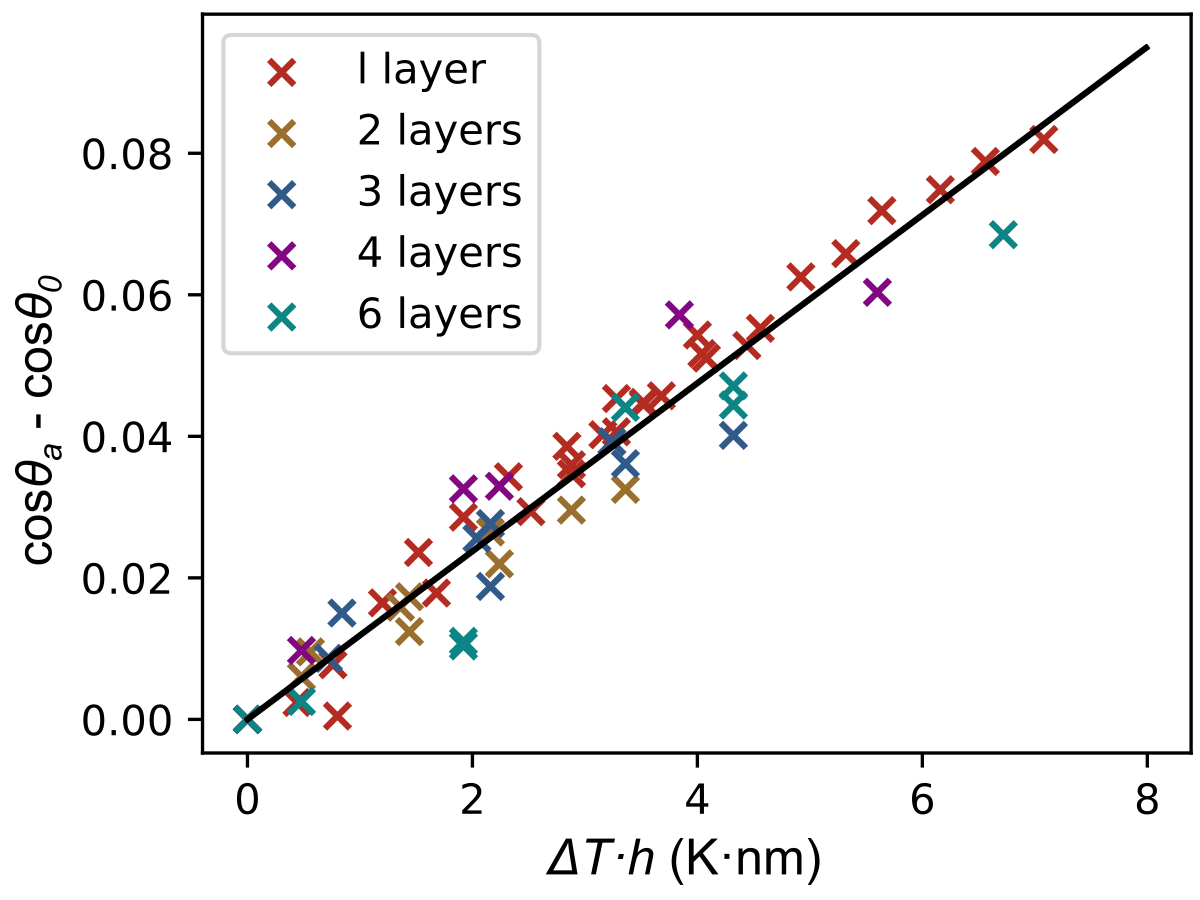}
  \caption{Collapse of the apparent contact angle data when plotted as $\cos\theta_a - \cos\theta_0$ versus the combined control parameter $\Delta T \cdot h$. Data from films of different thickness fall onto a single master curve, confirming that arrested melting is governed by the product $\Delta T \cdot h$, independent of droplet size.
  }
  \label{fig:drop-slope}
\end{figure}

We have observed in experiment that droplets confined within thicker solid films exhibit a significantly stronger morphological response to temperature changes than those in thinner films. Moreover, the slope of $\cos\theta_a$ versus $\Delta T$ increases systematically with the embedding depth, confirming that the relevant control parameter is the product $\Delta T \cdot h$, consistent with Eq.~\eqref{eqn:cos}. When plotted as a function of the combined control parameter $\Delta T \cdot h$, data obtained from films of different thickness collapse onto a single master curve (Fig.~\ref{fig:drop-slope}), confirming that $\Delta T \cdot h$ fully captures the thermodynamic control of the arrested melting.

A linear fitting of the collapsed data in Fig.~\ref{fig:drop-slope} results in $1.2 \times 10^7,\mathrm{K^{-1}m^{-1}}$, very close to the value $1.5\times 10^7,\mathrm{K^{-1}m^{-1}}$ calculated from $\Delta S_\text{fus} /\gamma _{lv}$, confirming the thermodynamic control. The height of the mono- and multi-layers is most probably smaller than estimated from the molecular all-trans length and atomic force microscopy measurements under room temperature, leading to the deviation.

In contrast, no dependence on droplet size is observed, in agreement with the scaling analysis. The linear and reversible dependence of $\theta_a$ on temperature demonstrates that the system remains close to thermodynamic equilibrium throughout the transition. Rather than completing the phase transition, the system continuously adjusts the droplet morphology to minimize the total free energy, resulting in a stable coexistence of liquid and solid well above the bulk melting temperature.

A direct consequence of interfacially arrested melting is an intrinsic amplification of temperature variations into macroscopic optical signals. In the suspended melting regime, small changes in temperature modify the balance between bulk melting enthalpy and interfacial energy, shifting the equilibrium apparent contact angle of the confined droplet. Because the droplet radius is much larger than the embedding depth, this shift corresponds to a large change in droplet profile and liquid–vapor interfacial area.

Experimentally, this amplification manifests as a pronounced displacement of Newton rings under minute temperature variations. As shown in Figure S1 in Supporting Information, temperature variations below $0.1^\circ$C produce clearly resolvable shifts in the interference pattern, indicating substantial morphological changes of the droplet despite minor bulk melting. The response is fully reversible and reproducible, reflecting an equilibrium adjustment rather than a kinetic process.

Importantly, this behavior does not rely on external transduction or material-specific sensing elements. Instead, the amplification arises directly from thermodynamic competition between bulk and interfacial energies under geometric confinement. The arrested melting regime therefore provides a general physical mechanism by which small thermal perturbations are converted into large, optically accessible responses, and suggests potential applications for designing high-resolution thermometry.

The suspended melting observed here is fundamentally distinct from nucleation hysteresis, the liquid and solid domains are both far larger than the critical nucleus size, the morphological response is fully reversible, and no time-dependent relaxation is detected within experimental resolution. These features indicate that the arrested state corresponds to a thermodynamically stable equilibrium rather than a metastable or kinetically limited configuration.

More generally, the mechanism identified here does not rely on material-specific chemistry but on geometric confinement and interfacial energy balance. Any thin film in which melting creates additional liquid–vapor or liquid–solid interfaces, while lateral advance is constrained, is expected to exhibit a similar competition between bulk and interfacial free energies. In this sense, interfacially arrested melting provides a general route to stabilizing partial phase transitions in confined geometries.

The present results complement earlier studies of superheating, surface freezing, and confined phase transitions. Unlike classical superheating, where melting is delayed by the absence of nucleation sites \cite{metois1989Pb111, jin1999melting}, the arrest reported here persists after nucleation and is controlled by interfacial energetics. This highlights a distinct thermodynamic pathway by which first-order phase transitions can be suspended in thin films.

We have identified an interfacially arrested melting regime in molecularly thin films, where liquid and solid phases coexist stably well above the bulk melting temperature. The arrested state is governed by a single thermodynamic control parameter, the product of the temperature offset and the embedding depth, and is independent of droplet size. This balance between bulk melting enthalpy and interfacial energy enables intrinsic thermodynamic amplification, converting small temperature variations into large, reversible morphological and optical responses. Owing to its interfacial origin, this mechanism is expected to be broadly relevant to first-order phase transitions in confined thin-film systems.

\section*{Supplementary Materials}
Further details on the experimental procedures, the energy analysis underlying the model, and an additional supporting figure are presented in the Supplementary Material.

\section*{Acknowledgement}
Discussion with H. M\"ohwald, R.-C. Mutihac and H. Chen are gratefully acknowledged. C. Jin acknowledges support from IMPRS on Biomimetic Systems during the conduct of the study.

\section*{Author Declarations}
\subsection{Conflict of interest}
The authors have no conflicts to disclose.
\subsection{Author Contributions}
C. Jin conceived the study. C. Jin and H. Riegler designed the experiments. C. Jin performed the experiments, analyzed the data, carried out the theoretical interpretation, and wrote the original draft of the manuscript. H. Riegler provided experimental resources. All authors discussed the results, reviewed the manuscript, and approved the final version.

\section*{Data availability}
The data that supports the findings of this study are available within the article as plots.  
Data for generating plots in this study are available from the corresponding author upon reasonable request.

\section*{References}

\bibliography{ref-susp.bib}
 
\end{document}